\documentclass[%
 reprint,
 amsmath,amssymb,
 aps,
prb,
]{revtex4-1}

\usepackage{graphicx}
\usepackage{dcolumn}
\usepackage{bm}

\begin{document}


\title{First-principle study of the melting temperature of MgO}

\author{Max Rang}
 \altaffiliation[Also at ]{Faculty of Science and Technology and MESA$^+$ Institute for Nanotechnology, University of Twente, P.O. Box 217, 7500 AE Enschede, The Netherlands}
\author{Georg Kresse}%
 \email{georg.kresse@univie.ac.at}
\affiliation{University of Vienna, Faculty of Physics and Center for Computational Materials Sciences, Sensengasse 8/12, 1090 Wien}

\date{\today}

\begin{abstract}
Using first-principles only, we calculate the melting point of MgO, also called periclase or magnesia. The random phase approximation (RPA) is used to include the exact exchange as well as local and non-local many-body correlation terms, in order to provide high accuracy. Using the free energy method, we obtain the melting temperature directly from the internal energies calculated with DFT. The free energy differences between the ensembles generated by the molecular dynamics simulations are calulated with thermodynamic integration or thermodynamic perturbation theory. The predicted melting temperature is $T_m^\text{RPA} =3043\pm 86$ K, and the values obtained with the PBE and SCAN functionals are $T_m^\text{PBE} = 2747 \pm 59$ K and $T_m^\text{SCAN} = 3032\pm 53 $ K.
\end{abstract}

\pacs{Valid PACS appear here}
\maketitle

\section{Introduction}
Magnesium oxide (MgO) is a textbook example of a strongly ionic solid in the rock salt structure. The strong bonds between the ions result in a very high melting temperature $T_m$ and stability, making MgO a prime refractory material. As a consequence of the high $T_m$, the experimental melting temperature has not been determined with high precision, with results ranging from 3040 (100) K \cite{Zerr1994} to 3250 (20) K \cite{Ronchi2001}. The most recent non-outlier of the experimental data puts the melting temperature at 3098 (42) K \cite{Dubrovinsky1997}.

\textit{Ab initio} melting temperatures of materials can be calculated with two types of methods. The \textit{free energy approach} \cite{CarSugino1995} considers the free energy difference between the solid phase and the liquid phase and approximates the $F(T)$ curve for both phases. The intersection between these curves then directly gives the melting temperature. The alternative method uses phase coexistence calculations and usually leads to the necessity for simulation cells with many hundreds, often even thousands of atoms \cite{morris1994melting,alfe2003Almeltingcoexist,pedersen2013direct,Hummel2013interfacepinning,morawietz2016van}. These are computationally quite demanding, especially for a strongly ionic substance like MgO, in part because of the relatively high plane-wave cutoff energy. The free energy approach including the RPA energy has recently succeeded in predicting the melting temperature of Si \cite{DornerSi} and in the present work we present a comparable calculation for MgO. The melting temperature of MgO has been calculated using DFT methods in the past \cite{DarioMgO,TangneyMgO}, albeit only in the local density approximation (LDA) and a generalized gradient approximation (GGA). Here we consider the Perdew-Burke-Ernzerhof (PBE) GGA\cite{Perdew_PBE_1996} and the more sophisticated strongly constrained and appropriately normed (SCAN) meta-GGA\cite{Sun_SCAN_2015} functional, as well as the RPA, using the low scaling RPA implementation in VASP \cite{VASP,KaltakRPA}.
\section{Theory}
The main ingredients of the present calculation are thermodynamic integration (TI) \cite{zwanzig1954rw} and thermodynamic perturbation theory (TPT). Both are methods to calculate the free energy difference of ensembles generated by two different Hamiltonians with potential energy terms $U_A\left(\mathbf{R}\right)$ and $U_B\left(\mathbf{R}\right)$ respectively. The free energy difference by TPT is
\begin{equation}
F_B - F_A = -\frac{1}{\beta} \ln\langle e^{-\beta \left(U_B(\mathbf{R}) - U_A(\mathbf{R})\right)}\rangle_{\lambda=0},
\end{equation}
where we will use the second order cumulant expansion \cite{zwanzig1954rw}, which minimizes the error while still giving an accurate free energy difference \cite{DornerSi}. The cumulant expansion is given by
\begin{equation}
\begin{split}
F_B - F_A & \approx \langle \Delta U \rangle_{\lambda=0} - \frac{\beta}{2} \langle \left(\Delta U - \langle \Delta U \rangle \right)^2\rangle_{\lambda=0},\hspace{5pt}  \\
 \Delta U & = U_B - U_A\, .
 \label{equ:deltaU}
 \end{split}
\end{equation}
The free energy difference given by TI is
\begin{equation}
F_B - F_A = \int_0^1 \langle U_B(\mathbf{R}) - U_A(\mathbf{R})\rangle_\lambda\, d\lambda\,.
\end{equation}
The notation $\langle A \rangle_\lambda$ implies the thermodynamic average of the quantity $A$ over an ensemble generated by a classical Hamiltonian for which the potential energy $U_\lambda(\mathbf{R})$ term is a mixture of $U_A$ and $U_B$:
\begin{equation*}
U_\lambda(\mathbf{R}) = (1-\lambda)U_A(\mathbf{R}) + \lambda U_B(\mathbf{R})\,.
\end{equation*}
This means that TI involves simulating an intermediate ensemble at various $\lambda$ points to compute the integral numerically, while TPT can be evaluated using only selected independent configurations of the $U_A$ ensemble. The accuracy of straight-forward numerical TI over a complicated function depends on the number of $\lambda$ points and more importantly, the convergence of the thermodynamic ensemble at each $\lambda$, while the accuracy of a free energy difference calculated by TPT is mostly dependent on the overlap in phase space between the ensembles of $U_A$ and $U_B$.

The path towards an accurate free energy is then clear. We must first calculate the free energy of some system for which the free energy is analytically known, in the case of the solid the harmonic crystal and in the case of the liquid the ideal gas. Using TI or TPT, one can then include the change of the interatomic interactions by calculating the change in the free energy when going to the DFT Hamiltonian. The complete process consists of a chain of thermodynamic integrals, reserving TPT for Hamiltonians for which the evaluation is too expensive to perform MD with. In practice, we only perform TPT to correct for $\mathbf{k}$-point sampling, or to integrate from DFT to RPA. In both cases, 
the ensembles connected by TPT  sufficiently overlap in phase space that  the variance
of the change of the potential energy $\Delta U$ is  small. Hence, TPT becomes very accurate.

The current work uses a fully \textit{ab initio} method for the liquid, i.e. we do not integrate from the ideal gas to some reference potential (e.g. Lennard-Jones), which has been common practice \cite{CarSugino1995,deWijs1998,Alfe2002}, but rather integrates directly. The numerical integral of the $U(\lambda)$ curve of this step is not easily evaluated, a problem which an intermediate reference potential would solve, but such an intermediate step can sacrifice the accuracy somewhat, especially since finding a reasonably accurate auxiliary potential for an ionic material such as MgO is not entirely trivial. More sophisticated methods of applying empirical potentials have recently been proposed \cite{FangZhu2017}, but the current work considers a simpler solution through coordinate transformation of the coupling constant integral from the ideal gas to the liquid, as also considered in \cite{DornerSi}.
\section{Methods}
With few exceptions, the liquid and solid were simulated in the same manner: NVT ensembles with a fixed number of atoms $N$, temperature $T$ and equilibrium volumes $V_{l(s)}$ were considered. The Nos\'e-Hoover thermostat \cite{nose,hoover} was used for the liquid phase, while for the solid phase, the Andersen thermostat \cite{andersen} was used. A Verlet algorithm implemented in VASP is used to perform the molecular dynamics through integration of Newton's equations of motion. The partial occupancies for the electronic calculation are calculated using a simple Fermi smearing with width $\sigma = k_B T$. The plane-wave cutoff was set to $400$ eV. The pseudopotentials used within the PAW formalism were the Mg\_pv and the O potentials. The Mg\_pv potential, which includes the semi-core $2p$ orbitals at the same level as the valence orbitals, is more expensive than the regular Mg potential, but also more accurate. Most importantly, the standard Mg potential possesses low-lying ghost states when another atom is at close range, which results in failure to converge.  The TI from the ideal gas to the liquid is thus not feasible using the standard potential. This is a problem common to virtually all elements with shallow semi-core states. Furthermore, the Mg $2p$ shell is highly polarizable and expected to yield a dispersion interaction with the oxygen $2p$ shell for the RPA. Its inclusion is therefore mandatory for high accuracy.

For the numerical thermodynamic integration, a simple Gaussian integration using 3 points was used in all cases, except for the integration from the ideal gas to the DFT liquid. There, to include the end points, a Gauss-Lobatto integration scheme using 8 points was used. The timestep was set at POTIM=$2.0$ fs for almost all simulations. The few exceptions are discussed below. The simulation times depend on the integration step, since integrating between two very similar Hamiltonians yields very quick convergence and requires a relatively short simulation time, while for dissimilar Hamiltonians more MD steps are necessary. In practice, we found that the integral from the ideal gas to the DFT liquid with electrons sampled at the $\Gamma$ point required many steps, typically 30000 MD steps per $\lambda$ point, to yield reasonable convergence, while an integral from the $\Gamma$ point to $2\times 2\times 2$ $\mathbf{k}$-points required only 8000 MD steps or less to yield satisfactory statistics.
\subsection{Liquid phase}
\begin{figure}
	\begin{center}
		\includegraphics[height=52mm,clip=true]{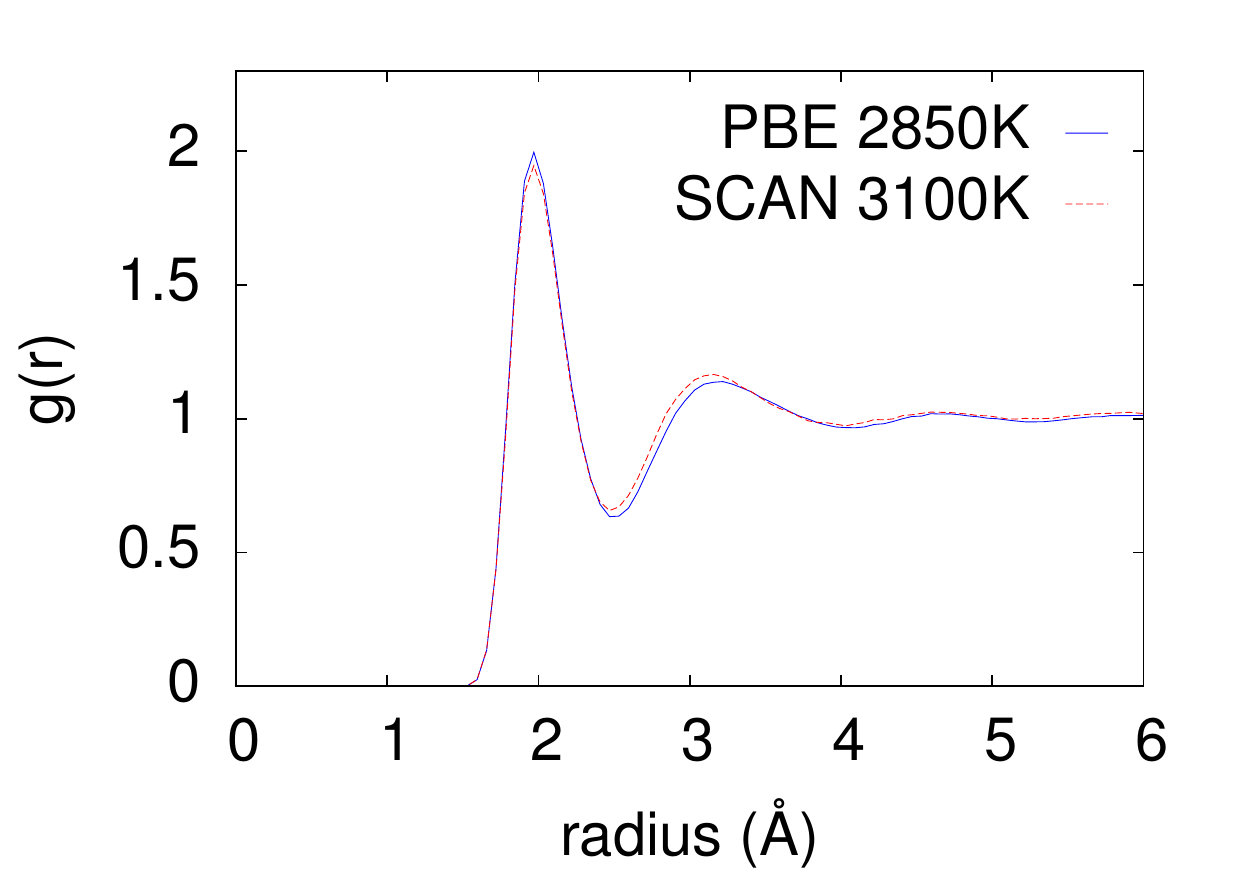}
	\end{center}
	\caption{
		Pair correlation functions of the liquid MgO at 2850 K for the PBE ensemble and 3100 K for the SCAN ensemble, respectively. In both cases, 128-atom unit cells were used.
	}
	\label{fig:pcdat}
\end{figure}
The starting point of the liquid phase is the diatomic ideal gas. The free energy of a diatomic free gas is just the sum of the free energies of the monoatomic free gas, as the particles do not interact. The expression for the free energy is
\begin{equation}
F_\text{ideal gas} = -k_B T \ln \left[ \frac{V^N}{\Lambda^{3N}N!}\right]\, ,
\end{equation}
where $\Lambda$ is the thermal de Broglie wavelength:
\begin{equation*}
	\Lambda = \frac{h}{\sqrt{2\pi m k_B T}}\, .
\end{equation*}
Calculating the free energy of the \textit{ab initio} liquid then becomes a matter of appropriately sampling the $\langle U(\lambda)\rangle_\lambda$ curve. This curve is not symmetric, even diverging for $\lambda \rightarrow 0$ because of Pauli and Coulomb repulsion and would require many $\lambda$ points to be accurately sampled in $\lambda$ space. This is computationally undesirable for obvious reasons. As in previous work, to circumvent this problem, a coordinate transformation is performed. Instead of over $\lambda \in \left[0,1\right]$, the integration is performed over $x\in \left[-1,1\right]$, where 
\begin{equation}
\lambda(x) = \left(\frac{x+1}{2}\right)^{\frac{1}{1-k}}\, .
\end{equation}
The integration then becomes
\begin{equation}
\int_0^1 f(\lambda)\,d\lambda = \frac{1}{2(1-k)} \int_{-1}^1 f(\lambda(x))\,\lambda(x)^k \, dx\, ,
\end{equation}
where we choose $k=0.73$. For very small $\lambda$ the ensemble is ideal gas-like, such that the atoms are allowed to come very close to each other. In order to ensure that the time integration with these interactions was sufficiently accurate, for these simulations the timestep was decreased to $0.5$ fs. Additionally, the Andersen thermostat \cite{andersen} with an interaction probability of 0.01 was used for these ensembles, since the ideal gas-like properties caused anomalies in the energies of the Nos\'e-Hoover thermostat, resulting in unreasonable behaviour in the instantaneous temperature.\\

\begin{figure}
	\begin{center}
		\includegraphics[width=\linewidth,clip=true]{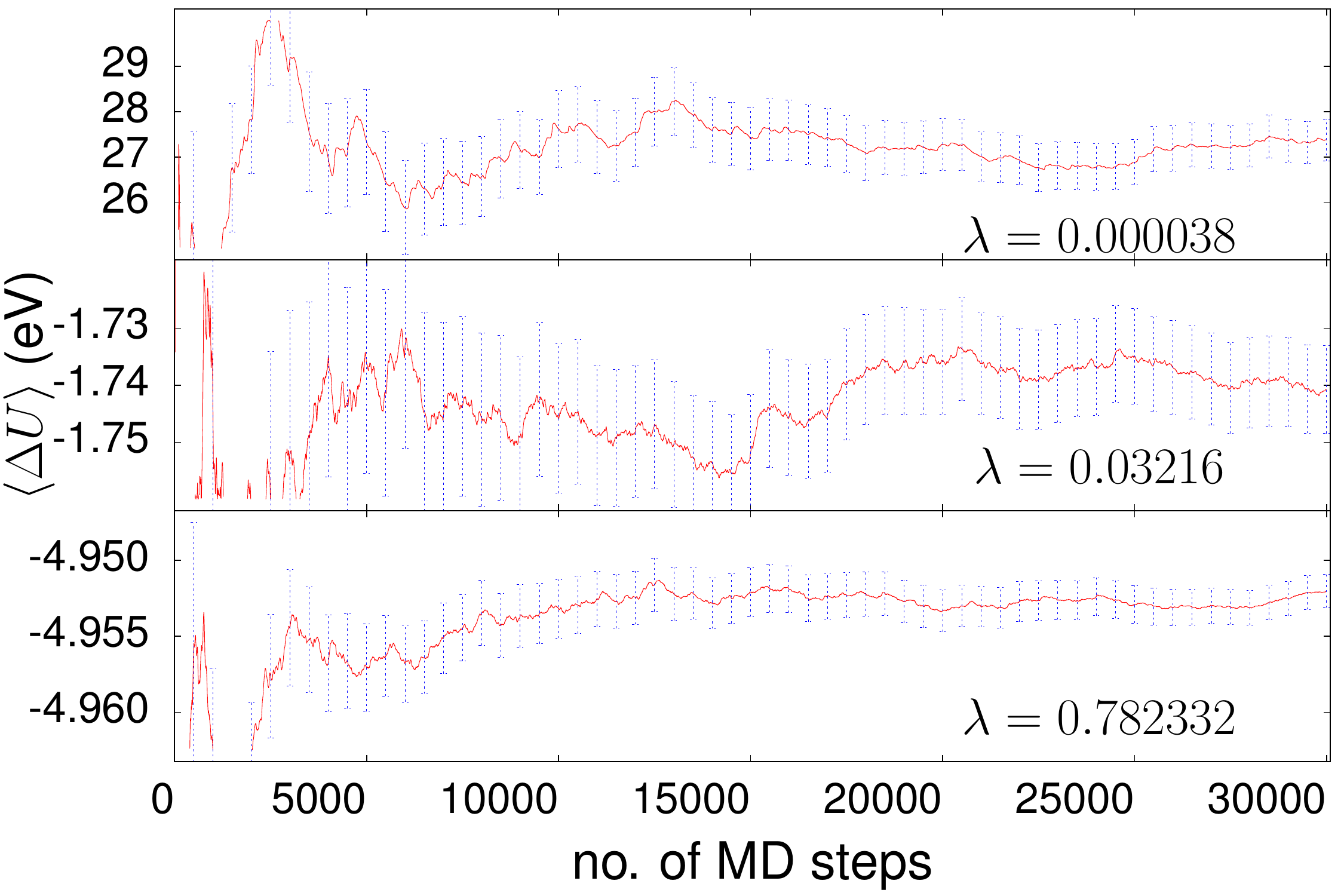}
	\end{center}
	\caption{
		The integrand of the TI from the ideal gas to the DFT Hamiltonian at various couplings $\lambda$, as a function of the number of MD steps. The errorbars are the block averaged standard deviations, plotted every 10 blocks. The error is by far largest for the smallest coupling, and the smallest for full coupling. In the first few thousand steps, the ensemble is equilibrating to the coupling constant, causing large fluctuations. These steps were excluded in the analysis.
	}
	\label{fig:lambda}
\end{figure}
Figure \ref{fig:pcdat} shows the pair correlation (PC) functions of the 128-atom liquid configurations. The fact that the PC functions for the PBE ensemble at 2850 K and the SCAN ensemble at 3100K overlap so closely shows that their thermodynamic states are very similar, such that if one of them is close to the melting temperature, so is the other. Most importantly, it suggests that the temperatures were chosen adequately for both functionals.\\

\begin{figure}
	\begin{center}
		\includegraphics[width=0.49\linewidth,clip=true]{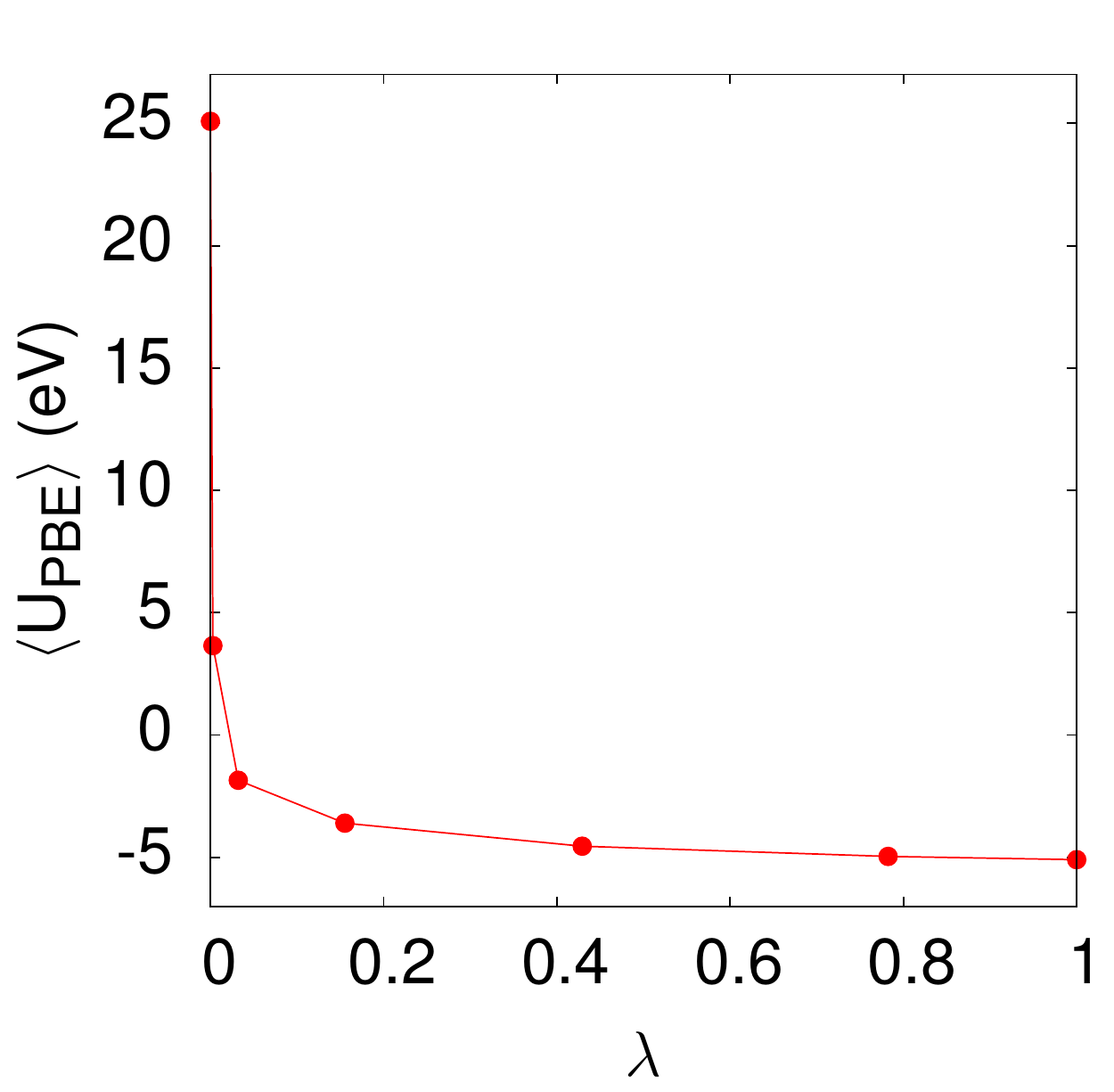}
		\includegraphics[width=0.49\linewidth,clip=true]{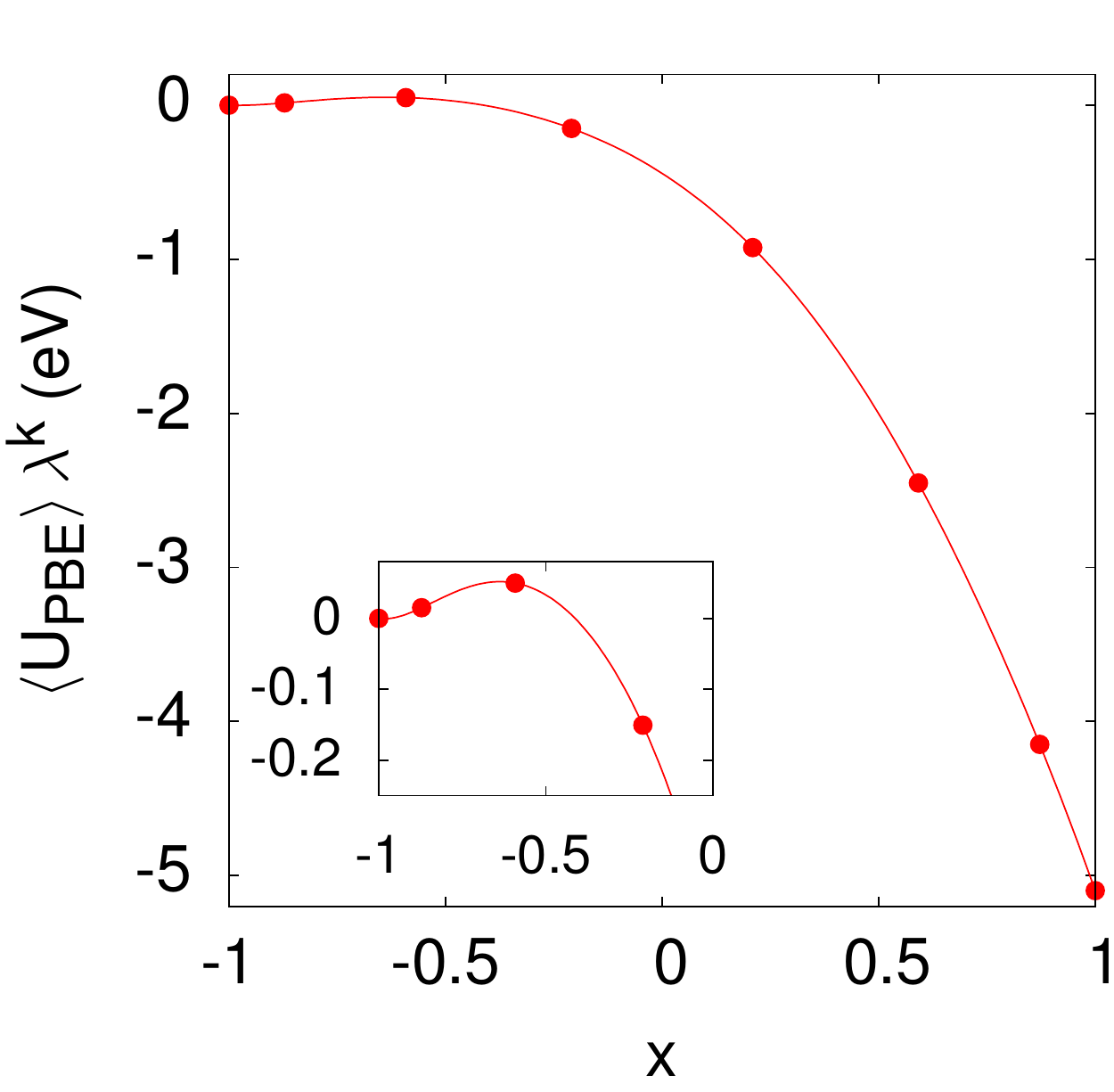}
	\end{center}
	\caption{
		Integrand of the TI from the ideal gas to liquid MgO using PBE 
		as a function of $\lambda$ and, after the transformation, as a function of $x$ for $k=0.73$.
		The inset in the right figure shows the negative $x$ range, corresponding to very small $\lambda$.
	}
	\label{fig:integral}
\end{figure}
The convergence of the integrand at various $\lambda$ values with respect to the simulation time is shown in figure \ref{fig:lambda}. For small $\lambda$, the variance can be very large, but the coordinate transform and the Gauss-Lobatto integration suppresses these contributions, such that the error in the free energy contribution from the full integration step is below 6 meV per atom, which results in a relative error of around $0.13 \%$.\\
 
Having obtained a free energy for the DFT liquid sampling the BZ only at the $\Gamma$ point, one can then improve the accuracy by considering more accurate Hamiltonians (i.e. more sophisticated exchange-correlation functionals, a denser $\mathbf{k}$-point grid, a higher plane-wave cutoff, etc.).
The individual steps used in the present work are summarizes in Tab.  \ref{tab:steps} and discussed in the next section.

\begin{table}[h]
	\caption{
		Individual integration steps used for the liquid and solid in the present work. ``DFT'' implies that the desired density
 functional is used. The ${\mathbf k}$-points are specified in parentheses.}
	\label{tab:steps}
	\begin{ruledtabular}
		\begin{tabular}{l}
			liquid MgO        \\                                                               
			\hline               
			TI ideal gas $\to$ PBE ($\Gamma$)	 	\\
			TI PBE ($\Gamma$)	 $\to$  DFT ($2 \times 2 \times 2$)           \\
			TPT from DFT ($2 \times 2 \times 2$) $\to$ RPA   ($2 \times 2 \times 2$)  \\
			\hline
			solid MgO         \\
			\hline               
			ideal crystal in supercell using  DFT ($2 \times 2 \times 2$) 			 \\
			harmonic	 contribution using  DFT ($2 \times 2 \times 2$)   \\
			TI harmonic $\to$  anharmonic  using  DFT ($2 \times 2 \times 2$)  \\
			TPT from DFT ($2 \times 2 \times 2$) $\to$ RPA  ($2 \times 2 \times 2$) 
		\end{tabular}
	\end{ruledtabular}
\end{table}

\subsection{Solid phase}
At T=0, the internal energy and free energy are equal, such that the exact free energy can be found directly using a DFT calculation for the ideal crystal. The contribution to the free energy from harmonic vibrations of the crystal (phonons) $F_\text{harm}$ can be described analytically as well
\begin{equation}
F_\text{harm} = \int_0^\infty D(\omega) \left[ \frac{\hbar \omega}{2} + \frac{1}{\beta} \ln \left(1-e^{-\beta \hbar \omega}\right)\right]d\omega.
\end{equation}
Here $D(\omega)$ is the phonon density of states. We calculate this using VASP and an auxiliary script, which calculates the dynamical matrix from the force constants written by VASP and diagonalizes it to find the phonon spectrum. Integrating the DOS is then straightforward. The second contribution to the free energy of the solid phase is the anharmonic contribution, which we find by performing TI from the harmonic crystal to the DFT solid.
\begin{equation}
	F_\text{anharm} = \int_0^1 \langle{U_\textit{DFT} - U_{\text{harm}}\rangle}_\lambda\,d\lambda
\end{equation}
Finally, through TI and TPT one can improve on this result by integrating to more accurate Hamiltonians, e.g. using a different functional, a denser $\mathbf{k}$-point grid, a higher cut-off energy, etc. The individual steps used in the present work are summarizes in Tab.  \ref{tab:steps}  and discussed in the next section.

\section{Results}
The melting temperature is evaluated as
\begin{equation}
T_m = T - \frac{F_l -F_s}{S_l - S_s}
\label{eq:Tm}
\end{equation}
where the entropy $S_{l,s}$ is calculated as $S_{l,s}=(U_{l,s}-F_{l,s})/T$ and the subscript $s$ ($l$) denote the solid (liquid) phase at $T$. This equation is valid in the range where the Helmholtz free energy $F_{l,s}$ is linear in the temperature, so the difference in the working temperature $T$ and the calculated melting temperature $T_m$ must not be too large. In our final result, the temperatures are assumed to be close enough that linearization is allowed. In the above we have assumed zero pressure, so the equilibrium volume for each functional must first be determined. The equilibrium volume can be found by requiring that the trace of the macroscopic stress tensor $\bar{\sigma}$ \cite{nielsen1983first} which is defined as (in the usual units energy per volume or force per square area)
\begin{equation*}
		\bar{\sigma} = \frac{1}{V} \Big( \Big\langle \frac{\partial U}{\partial \epsilon} \Big\rangle + {\rm diag}(N k_B \langle T\rangle)\Big)\, ,
\end{equation*}
is zero. Here $\epsilon$ is the strain tensor, and ${\rm diag}$ implies a diagonal matrix.
To that end, an NVT ensemble was simulated at various volumes $V$, to obtain a pressure-volume curve yielding directly the equilibrium volume by intersection with the $x$ axis.

For the liquid, the actual pathway is the following. First we set the volume to either the PBE or SCAN equilibrium volume as previously determined and calculate the free energy of the ideal gas. Then we integrate to the DFT Hamiltonian using the PBE functional sampling the BZ at the $\Gamma$ point only. From there,  we either integrate to the PBE Hamiltonian with $2\times 2 \times 2$ $\mathbf{k}$-points, or to the SCAN Hamiltonian with $2 \times 2 \times 2$ $\mathbf{k}$-points (compare
Tab. \ref{tab:steps}). 

For the solid, we set the volume to the PBE or SCAN equilibrium volume and calculate the ideal crystal internal energy using the desired supercell (128 or 64 atoms) and $2 \times 2 \times 2$ $\mathbf{k}$-points. The force constants are calculated using the same supercell and finite differences for both PBE and SCAN at the previously determined volumes again employing $2 \times 2 \times 2$ $\mathbf{k}$-points. The harmonic free energy is calculated from the phonon density of states. Finally, the anharmonic contributions are determined by integrating from the
harmonic to the full DFT Hamiltonian (PBE or SCAN) (also compare Tab. \ref{tab:steps}). We note that any errors in the harmonic force constants,
for instance caused by noisy forces, will drop out upon integration to the DFT Hamiltonian.
 
\begin{figure}[h]
	\begin{center}
		\includegraphics[height=50mm,clip=true]{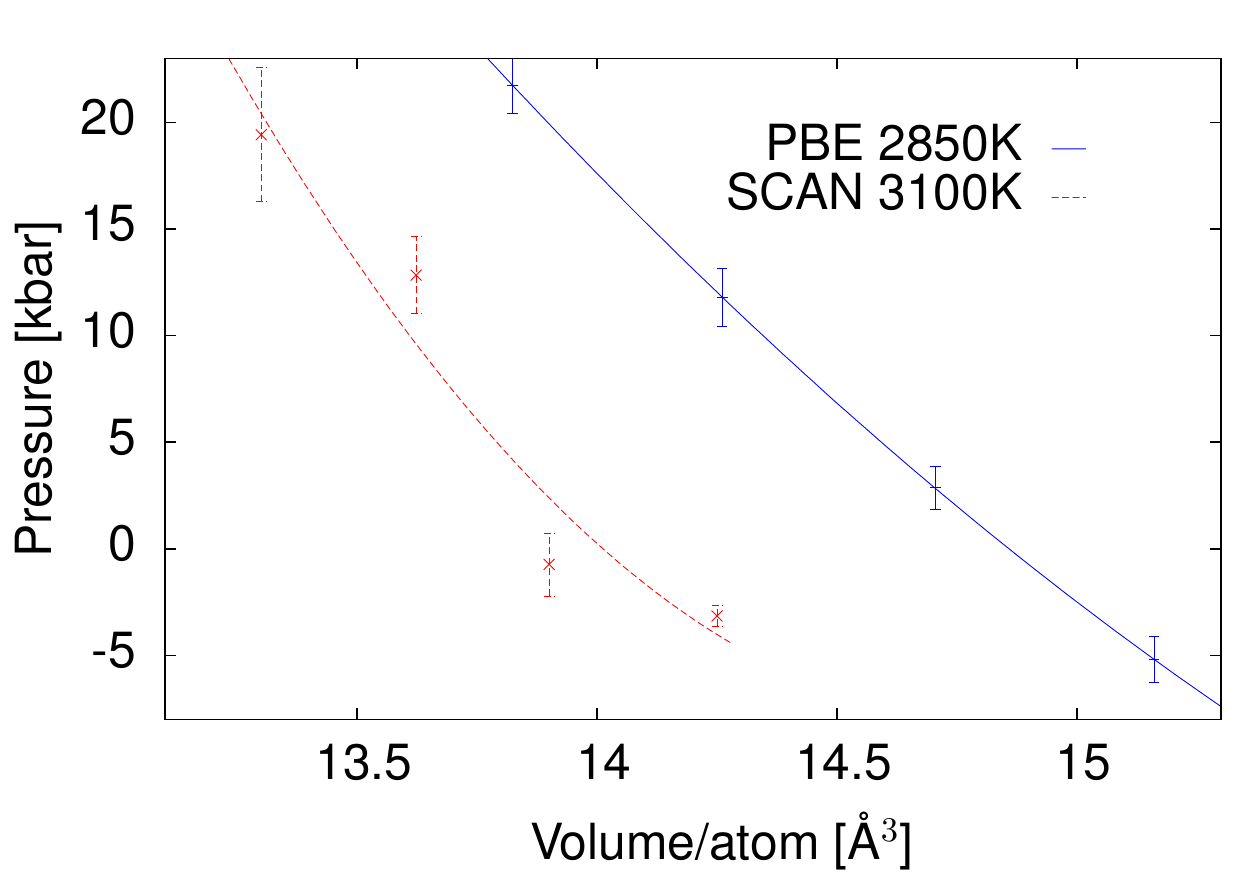}
		\includegraphics[height=50mm,clip=true]{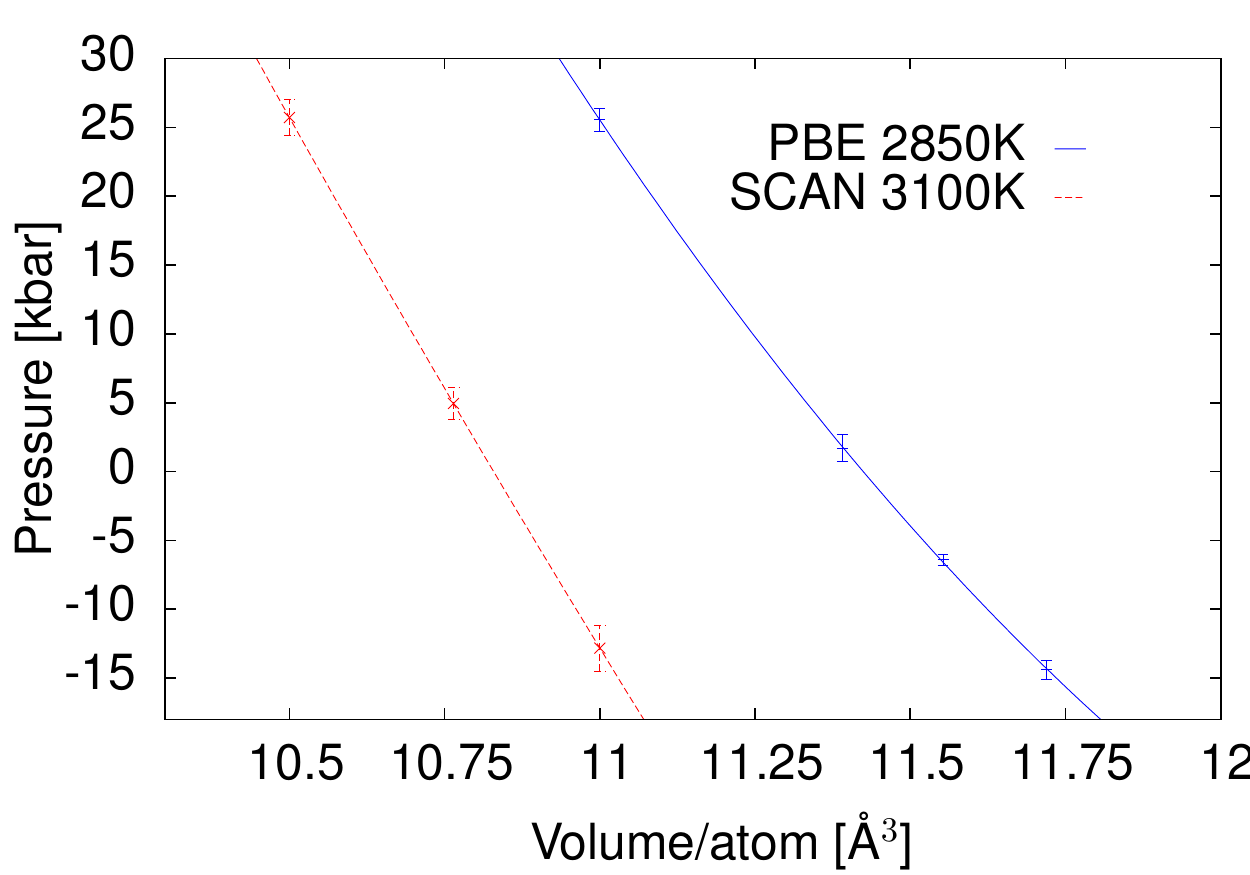}
	\end{center}
	\caption{
		The pressure of liquid and solid phases of MgO as a function of the volume for particular functionals and temperatures. The pressures have been converged with respect to the plane wave cutoff by including a correction commonly referred to  as Pulay stress.
	}
	\label{fig:PV}
\end{figure}

\begin{table}[h]
	\caption{
		Equilibrium volumes for both PBE and SCAN functionals, as well as the free energy differences calculated by TI of each step. The melting temperature, entropy of fusion and melting slope are also reported for each functional.
	}
	\label{tab:summary}
	\begin{ruledtabular}
		\begin{tabular}{lrrr}
			DFT functional           &  PBE      &   SCAN     &  SCAN   \\
			Number of atoms			 &  128      &   128      &   64    \\
			T(K)                     & 2850 K    &  3100 K    &  3100 K \\
			\hline
			solid MgO         \\
			volume 	                 &11.39 \AA$^3$   &  10.76  \AA$^3$  &  10.76  \AA$^3$  \\
			crystal at T=0			 &-5.8294       &  -8.5375	&-8.5364	 \\
			$\Delta F$ harmonic		 &-1.4898      &  -1.3970  & -1.3945   \\
			$\Delta F$ anharmonic    &0.0528      &  -0.1275  & -0.1288     \\
			\hline                                                            total $F_s$  & -7.2664 eV  &  -10.0621 eV &  -10.0598 eV         \\      
			\hline               
			liquid MgO       &&& \\                                                               
			volume                   &  14.87 \AA$^3$   &  13.62 \AA$^3$    &  13.62 \AA$^3$  \\
			ideal gas				 &-2.9890   	&  -3.2669 	& -3.2573	\\
			PBE $\Gamma$			 &-4.2894       &  -4.2287  & -4.2344       \\
			DFT $2 \times 2 \times 2$&-0.0024       &  -2.5765  & -2.5780        \\
			\hline
			total $F_l$ & -7.2808 eV & -10.0721 eV & -10.0697 eV\\
			\hline
			$T_m$                    & $2747\pm 59$ K &  $3032\pm53$  K  & $3031\pm72$ K    \\
			$\Delta s$				 & $1.631  $ k$_B$&  $1.699  $ k$_B$  & 1.660 k$_B$     \\
			$dT_m/dP$				     & 155 K/GPa   &   134 K/GPa  & 138 K/GPa      
		\end{tabular}
	\end{ruledtabular}
\end{table}

Table \ref{tab:summary} shows the results of the calculations using a 128-atom ensemble.  We note that the contribution labeled as `DFT $2\times 2 \times 2$' for the liquid  is so large for SCAN compared to PBE, because we integrate from the PBE functional to the SCAN functional in this step, yielding a fairly large mostly constant offset related to the change of the exchange correlation functional.  The melting temperature as calculated by PBE is around $400$ K too low at $2747\pm 59$ K. This underestimation is in line
with the underestimation for another GGA calculation published in previous work, where  a melting temperature of $2550\pm100$~K was predicted \cite{DarioMgO} (mean of values reported in abstract and tables).
The SCAN functional slightly underestimates the melting temperature at $3032\pm 53$ K, but is very close to the range of experimental values, which is $3040 - 3250$ K. 
Previous calculations indicate that the local density approximation also yields quite accurate melting temperatures around $3110\pm50$~K \cite{DarioMgO}.

Since a 128-atom ensemble is somewhat large to perform many RPA calculations, we also considered a 64-atom ensemble.
The agreement between the calculations using 64 atoms and 128 atoms is remarkably good, in line with our previous observations
for silicon that the predicted melting points hardly depend on the  considered system size\cite{DornerSi}. 
We note that the ${\mathbf k}$-point sampling was not increased from 128 to 64 atoms, although this would be 
required to maintain a constant ${\mathbf k}$-point sampling error. In fact, the absolute energies per atom are slightly shifted for both
the solid and the liquid. That the shift is roughly the same for both phases relates to  the liquid and solid possessing a band gap, 
resulting in a similar ${\mathbf k}$-point sampling error. Concerning the statistical errors, we note that the simulation length was roughly
the same for the 128 and 64 atom ensembles. This causes an increased statistical error for the 64 atom ensemble, as the variance decreases with system size as $1/\sqrt{N}$.

To determine the melting point on the level of the random phase approximation, we  integrated from SCAN to the  RPA using the 64 atom
ensembles.  We determined the RPA correlation energy for 20 (60) solid (liquid) independent configurations and performed TPT to calculate the free energy difference. 
The variance of the change of the energy $\Delta U$ (Equ. \ref{equ:deltaU}) is fairly small indicating that the SCAN and RPA ensembles are very similar.
This is also confirmed by the small change in the calculated melting temperature. The melting temperature predicted by the RPA is $T_m^\text{RPA} = 3043\pm 86$ K. The entropy of fusion as well as the derivative of the melting temperature with respect to the pressure also do not change much and  are $\Delta s = 1.724$ k$_B$ and $dT_m/dP = 132$ K/GPa, respectively.

\section{Summary and Conclusions}
We have calculated the melting point of MgO, a diatomic highly ionic compound, using the free energy approach and the random phase approximation. The predicted melting temperature $T_m^\text{RPA} = 3043\pm 86$ K at zero pressure is in very good agreement with the lowest bound of the experimental values \cite{Zerr1994}. It must be mentioned though that the experimental values vary wildly, with the highest experimental value \cite{Dubrovinsky1997} being approximately 200 K higher than our calculated melting temperature. The SCAN functional predicts a melting temperature of $T_m^\text{SCAN} = 3032\pm 53$ K, nearly equal to the RPA (and close to previous melting temperature predictions for the LDA). 
The PBE melting temperature is too low at $T_m^\text{PBE} = 2747 \pm 59 $ K,
in line with previous studies that also predicted a too low GGA melting temperature.  We found that using modest ensemble sizes of around 64 atoms yields good agreement with the calculation for many atoms (in this case 128 atoms), keeping the computational requirements relatively low, especially compared to the ensemble size requirements of the alternative coexistence methods. The methodology translates easily to different materials as well as non-zero pressures, such that through this method, the full phase diagram of a material can be calculated  from first principles even beyond density functional theory. 

In summary, after Si, this is now the second system for which the RPA yields a very good to excellent description of the melting temperature. This substantiates our hope that
quantitative predictions of melting temperatures are now within reach for many more materials.  It is particularly noteworthy that the RPA seems
to work equally well for Si and MgO. With the limited experience we yet have, we can furthermore state that the SCAN functional also yields reasonable
melting temperatures, certainly more concise than those for the local density approximation (huge errors for Si) or generalized gradients approximations (large errors for MgO). 
We will come back to a more extensive review of different functionals in future work.

\bibliography{literature}

\end{document}